\documentclass[%
 aip,
 amsmath,amssymb,
 reprint,%
]{revtex4-1}

\usepackage{graphicx}
\usepackage{dcolumn}
\usepackage{bm}
\usepackage{siunitx}
\usepackage[utf8]{inputenc}
\usepackage{lineno}
\usepackage[T1]{fontenc}
\usepackage{mathptmx}
\usepackage{etoolbox}
\usepackage{verbatim}
\makeatletter
\def\@email#1#2{%
 \endgroup
 \patchcmd{\titleblock@produce}
  {\frontmatter@RRAPformat}
  {\frontmatter@RRAPformat{\produce@RRAP{*#1\href{mailto:#2}{#2}}}\frontmatter@RRAPformat}
  {}{}
}%
\makeatother
\begin{document}

\preprint{AIP/123-QED}

\title{Realization of independent contacts in barrier-separated InAs/GaSb quantum wells}
\author{Xingjun Wu}
\thanks{Authors to whom correspondence should be addressed: [Xingjun Wu, wuxj@baqis.ac.cn; Rui-Rui Du, rrd@rice.edu] }
 \affiliation{Beijing Academy of Quantum Information Sciences, Beijing, 100193, China}
 
 \author{Jianhuan Wang}
 \affiliation{Beijing Academy of Quantum Information Sciences, Beijing, 100193, China}

\author{Miaoling Huang}
 \affiliation{Beijing Academy of Quantum Information Sciences, Beijing, 100193, China}
 
\author{Shili Yan}
 \affiliation{Beijing Academy of Quantum Information Sciences, Beijing, 100193, China}

\author{Rui-Rui Du}
\thanks{Authors to whom correspondence should be addressed: [Xingjun Wu, wuxj@baqis.ac.cn; Rui-Rui Du, rrd@rice.edu] }
\affiliation{%
International Center for Quantum Materials, School of Physics, Peking University, Beijing 100871, China}

\date{\today}

\begin{abstract}
InAs/GaSb double quantum wells (QWs) separated by a 100 \AA\ AlSb middle barrier are grown by molecular beam epitaxy. We report a nanofabrication technique that utilizes the surface Fermi level pinning position in InAs $[E_f^s(\rm InAs)]$ for realizing independent electric contacts to each well. In particular, separate ohmic contacts to the upper InAs quantum well are achieved by selectively etching down to the InAs, while contacts to the lower GaSb quantum well are obtained by the depletion method. For the latter, the upper InAs quantum well is locally pinched off by top etched trenches capped with a remaining 2-3 nm InAs layer. As a result of a relatively low $E_f^s(\rm InAs)$, applying a negative bias gate potential will create a conducting hole channel in GaSb, and hence a separate ohmic contact to the lower quantum well. This method is demonstrated with experiment and the support of a self-consistent band bending calculation. A number of experiments on separately probing Coulomb and tunnel-coupled InAs/GaSb systems now become accessible.

\end{abstract}

\maketitle

Layer index is an important degree of freedom in low-dimensional correlated electronic systems. A wide variety of exotic layer-layer correlated quantum phases have been reported in recent decades, such as Bose-Einstein condensation of excitons in GaAs or graphene bilayers\cite{eisenstein2004bose,nandi2012exciton,tutuc2004counterflow,liu2017quantum,li2017excitonic}, the correlated insulating states and superconductivity in moir\'e-coupled graphene superlattices\cite{cao2018correlated,cao2018unconventional,serlin2020intrinsic,chen2020tunable,liu2020tunable,chen2021electrically,cao2020tunable}, and recently the Mott insulator state and various generalized Wigner crystal states in transition metal dichalcogenides moir\'e superlattices\cite{regan2020mott,chu2020nanoscale,xu2020correlated,huang2021correlated}. Experimentally, transport studies based on independent electric contacts,  e.g., resonant interlayer tunneling\cite{spielman2000resonantly,yoon2010interlayer,burg2018strongly}, counterflow transport\cite{tutuc2004counterflow,li2017excitonic}, and Coulomb drag measurement\cite{nandi2012exciton,gorbachev2012strong,liu2017quantum,li2017excitonic}, have many prominent advantages compared with "parallel" studies and hence are widely used in these closely spaced bilayer or mulitlayer systems. 

In GaAs/AlGaAs double QW structures, by using precisely controlled etching, shallow ohmic contact\cite{patel1994independent}, or selective depletion of quantum wells underneath the top and bottom Schottky gates\cite{eisenstein1990independently}, independent contacts have been realized.  However, for GaAs/AlGaAs systems, it is difficult to host both two-dimensional electron gas (2DEG) and hole gas (2DHG) in closely spaced double QWs due to the large band gap of 1.5 eV in GaAs. In contrast, due to a unique inverted band structure, InAs/GaSb QW naturally hosts spatially closed 2DEG and 2DHG in the growth direction. Many exotic interlayer correlated states are predicted theoretically in this system, e.g., an exciton Bose-Einstein condensate, topological excitonic insulator, or a possible charge density wave state, which have attracted recent experimental attention \cite{lyo2015excitons,datta1985possibility,naveh1996excitonic,pikulin2014interplay,xue2018time,du2017evidence,wu2019electrically,wu2019resistive}. Apart from the charge carriers of opposite polarity in closely spaced InAs and GaSb wells, it is also of interest due to the topological properties of the system. Tunnel-coupling between wells leads to the opening of a hybridization gap hosting topologically protected helical edge states, making InAs/GaSb QW system a quantum spin Hall(QSH) insulator\cite{liu2008quantum,knez2011evidence,qu2015electric,couedo2016single,mueller2015nonlocal,du2017tuning,du2015robust}. The interplay of the quantum spin Hall effect and interlayer correlation makes this system more fascinating\cite{pikulin2014interplay,xue2018time,du2017evidence,wu2019electrically}. However, up to now, there is still a lack of transport experiments based on independent electric contacts to study InAs/GaSb structures. The major problem arises from the difficulty of fabricating independent ohmic contacts to InAs/GaSb to probe the systems separately. Some methods available for GaAs material systems, e.g., selective depletion of quantum wells by top and bottom gates, become invalid for InAs/GaSb. In this Letter, we demonstrate a viable nanofabrication method to provide ohmic contacts to the individual layers in the barrier-separated InAs/GaSb QW system. 
 
\begin{figure*}[htb]
	\begin{center}
		\includegraphics[width=0.8\linewidth]{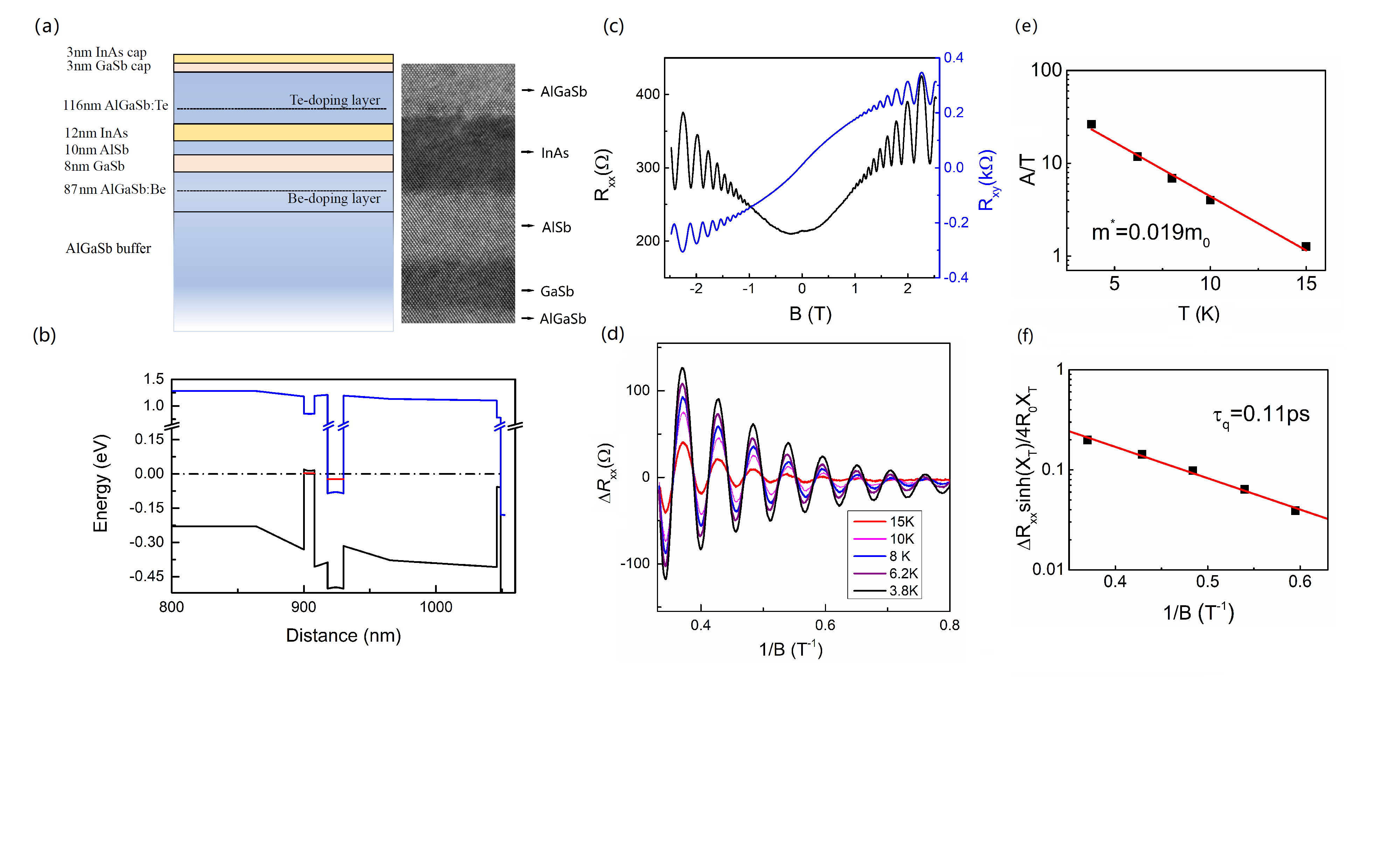}
	\end{center}
    \caption{ (a) Sketch of barrier-separated InAs/GaSb double quantum wells and the corresponding cross-sectional TEM image. (b) The self-consistent band structure calculation in the epitaxial region. (c) The magneto-resistance data for Hall bars when both QWs are contacted at 0.3 K. (d) The oscillatory longitudinal resistance $\Delta R_{xx}$, determined by subtraction of a parabolic fit to the background resistance, plotted versus $1/B$ for different temperatures $T$. (e) shows the Dingle plot of the SdH oscillation amplitude A divided by $T$ versus the temperature variance $T$. (f) shows the Dingle plot of $\Delta R_{xx}sinh(X_T)/4R_0X_T$ versus 1/B, data points are obtained by the magneto-resistance curve at $T$ = 3.8 K.}
    \label{fig:1}
\end{figure*}

Figure~\ref{fig:1}
(a) shows a schematic cross section of the InAs/GaSb structure. 12 nm-InAs/8 nm-GaSb QWs separated by 10 nm thickness AlSb middle barrier are sandwiched between $\rm Al_{0.8}Ga_{0.2}Sb$ barriers. Te and Be modulation doping layers of 1nm and 2nm thicknesses are positioned at 35 nm away from InAs and GaSb QWs, respectively, with concentrations of $1\times 10^{18}$ $\rm cm^{-3}$.  The quantum wells are grown in Veeco Gen III molecular beam epitaxy system on an InAs (100) substrate, which is equipped with valved cracker sources for group V (Sb2 and As2) fluxes and Knudsen cells for Ga/In/Al. An AlSb nucleation layer and a 900 nm  $\rm Al_{0.8}Ga_{0.2}Sb$ buffer are grown before the growth of the 87 nm $\rm Al_{0.8}Ga_{0.2}Sb$ lower barrier and the active QW region. All layers are grown at 500 \SI{}{\degreeCelsius}. The cross-sectional high-resolution TEM image of the sample is shown in Fig. 1(a). We find that the crystalline structure remains coherent across quantum well interfaces. More details about the material quality metrics can be found in the supplementary material. In this sample, the low-temperature electron density at zero bias is $8.1\times 10^{11}$ $\rm cm^{-2}$ with a mobility of $\sim$  $1 \times 10^{5}$ $\rm cm^{2}V^{-1}s^{-1}$, and the hole density at zero bias is $7.7\times 10^{11}$ $\rm cm^{-2}$ with a lower mobility of $\sim$ $1 \times 10^{4}$ $\rm cm^{2}V^{-1}s^{-1}$.  The density of the higher mobility carrier (electrons in this case) can be readily determined from the periodic oscillations of Shubnikov-de Haas (SdH) in magnetic fields in Fig.~\ref{fig:1}(c). The hole density can be deduced by sweeping back gate bias at a fixed magnetic field (see the supplementary material). The other structure used is similar to this one, but with 9.5 nm-InAs/10 nm-AlSb/5 nm-$\rm In_{0.25}Ga_{0.75}Sb$ QWs instead(not shown), and we call them sample A and B, respectively. Our method described below works for both structures. Figure~\ref{fig:1}(b) plots the self-consistent band structure for sample A. Here we assume the energy level of Te(Be) is 20 meV below(above) the conduction(valence) band edge of AlGaSb. Simulations of the band structure are carried out in Nextnano software by solving 1D Schr$\ddot o$dinger and Poission equations self-consistently\cite{birner2007nextnano}. Due to a type-II band alignment, with the conduction band minimum located in InAs and the valence band maximum in GaSb, the first subbands in electron and hole wells are occupied simultaneously. The calculation result is supported by low-temperature magneto-transport data, as seen in Figure~\ref{fig:1}(c). We observe prominent two-carrier transport behaviors in Hall-bar devices: 1) The longitudinal resistance $R_{xx}$ shows a positive magnetoresistance. 2) The Hall resistance $R_{xy}$ is nonlinear and becomes non-quantized in the high field region. 

\begin{figure}[tb]
	\begin{center}
		\includegraphics[width=1\linewidth]{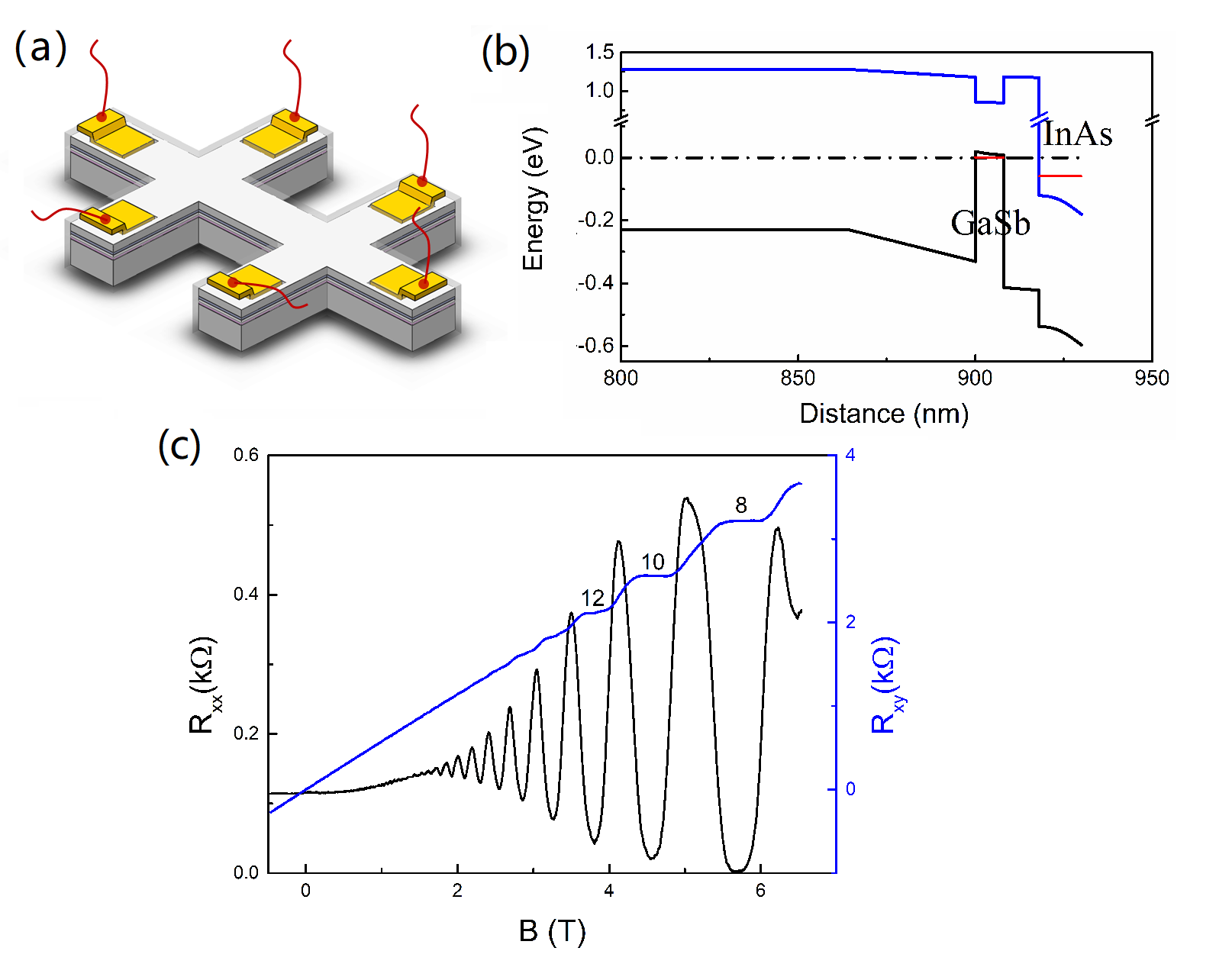}
	\end{center}
    \caption{ Independent contact for InAs QW. (a) and (b) Sketch of the device with independent contacts for InAs QW and the corresponding self-consistent band structure calculation. (c) The magneto-resistance data for Hall bars when only InAs QWs are contacted at 0.3 K. }
    \label{fig:2}
\end{figure}

As is well known, the SdH oscillations in small magnetic field can be described by the following equation:
$\Delta R_{xx}=4R_0[X_T/sinh(X_T)]exp(-D_q/\hbar w_c)cos(2\pi E_F/\hbar w_c-\pi)$,
where $R_0$ is the zero field resistivity, $D_q=\pi\hbar/\tau_q$ is the Dingle damping parameter, and $X_T=2\pi^2k_BT/\hbar w_c$ with $k_B$ being Boltzman constant and $\hbar w_c$ is the cyclotron energy. Here, the SdH oscillations are analyzed by fitting a parabolic function as a smooth magnetoresistance background to extract the oscillating component $\Delta R_{xx}$. The component $\Delta R_{xx}$ is plotted versus $1/B$ at different temperatures in Fig.~\ref{fig:1}(d). The effective mass $m^* = 0.019m_0$ can thus be extracted from the Dingle plot, where $m_0$ is the free electron mass. This value is consistent with that $(m^* = 0.023m_0)$ reported  previously in InAs\cite{nakwaski1995effective}, indicating the coupling from the adjacent GaSb layer positioned at 10nm away is insignificant. It is also evidenced by the transport result of the device made of independent contacts to InAs, as shown in Fig.~\ref{fig:2}.
In addition, by extracting from the Dingle plot in Fig.~\ref{fig:1}(f) and the mobility, we obtain the quantum lifetime ($\tau_q$ = 0.11 ps) and the transport lifetime ($\tau_t$ = 1.1 ps), respectively. We find that the two relaxation times are improved compared to that reported in a wider InAs/AlSb QW grown on GaAs substrate\cite{shojaei2015studies}. The improved transport properties are probably attributed to the lattice-matched substrate with fewer dislocations and smaller interface roughness. By comparing the lifetime ratio of this sample ($\tau_t/\tau_q$ = 10) with that of the high-mobility GaAs/AlGaAs samples\cite{sarma1985single}, we infer that short-range scattering, such as interface roughness and charged dislocation, still may dominate the scattering process.

In the following, we first describe the nanofabrication process of realizing an individual contact to InAs QW in our sample. Figure ~\ref{fig:2}(a) shows a sketch of the device. After the standard Hall-bar mesa is defined using wet chemical etching with an $\rm H_3PO_4/H_2O_2/C_6H_8O_7/H_2O$ solution, selective wet chemical etchants of $\rm C_6H_8O_7/H_2O_2$ and $\rm NH_4OH/H_2O$ are used sequentially to remove InAs cap and AlGaSb up-barrier at six ends of the leads in the Hall-bar device. The buried InAs QW is thus exposed to the surface. Subsequently, a 50 nm-thick Ti/Au layer for Ohmic contacts to InAs is deposited without annealing. In order to prevent the TiAu electrodes from attaching to 2DEG or 2DHG through the sidewall, a dielectric layer of $\rm Al_2O_3$ is deposited before TiAu evaporates.  Fig.~\ref{fig:2}(b) gives the band bending calculation results for the device. It can be found that simply by selectively etching down to the InAs layer, the surface Fermi level pinning position is still located above the first subband of InAs QW, making metallic electrodes that subsequently deposited easy to connect to the 2DEG in InAs. The result of low-temperature magneto-transport measurement verifies that this scheme is highly effective for independent contacts to the InAs layer.  As shown in Fig.~\ref{fig:2}(c), we observe standard SdH oscillations accompanied by well-quantized Hall resistances $(R_{xy})$. Moreover, the carrier density extracted from the SdH oscillation is consistent with that from low-field Hall measurement. In contrast to Fig.~\ref{fig:1}(c), it is reasonably concluded that the transport is predominately from the InAs layer.

In order to individually contact to GaSb QW, a natural idea is to deposit metal electrodes to the lower surface of GaSb QW. To this end, a "flip-chip" fabrication is necessary\cite{weckwerth1996epoxy}.  This technique requires not only difficult thinning of the sample from the bottom but also precise control of etching to the GaSb QW surface. Since a lack of wet chemical etchants with high selectivity of AlGaSb over GaSb, precise control of etching down to the GaSb QW becomes difficult. Moreover, since GaSb material doesn't posses surface states like InAs, the metal/GaSb interface is prone to forming a Schottky barrier, making it difficult to produce reliable low-resistance ohmic contacts. We circumvent this problem by using the quantum well depletion method to pinch off the upper InAs QW. Previously, in closely spaced 2DEGs or 2DHGs of GaAs, the depletion method is achieved either by applying top and bottom Schottky gates\cite{eisenstein1990independently} or selective etching of modulation-doped layers\cite{lang2008selective}. However, the methods become unavailable for InAs/GaSb structures: 1) For the former, charge compensation effect will be produced when Schottky gates are applied in closely spaced electron-hole double layers, i.e., while electrons in InAs are being depleted, holes will be accumulating in GaSb due to the imperfect screening, making InAs conducting channel difficult to be pinched off; 2) For the latter, the upper and lower modulation-doped barriers in InAs/GaSb consist of the same AlGaSb materials, making selective etching impossible.  

\begin{figure}[!tb]
	\begin{center}
		\includegraphics[width=1\linewidth]{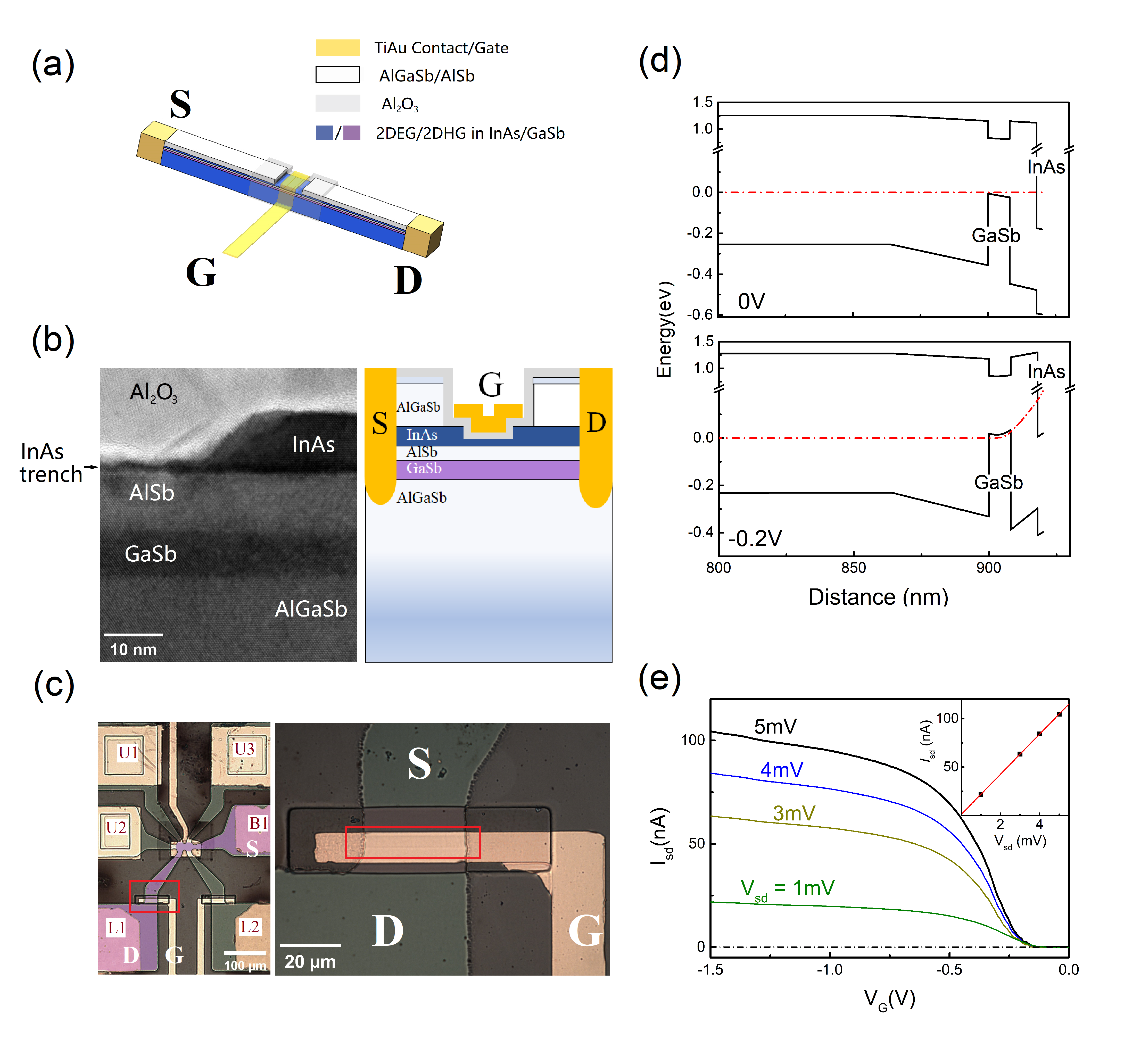}
	\end{center}
    \caption{ Independent contact for GaSb QW. (a) Sketch of the studied channel in the device. (b) The cross section diagram and the corresponding TEM image for independent contacts to GaSb QW. (c) The optical image of the device and the zoomed image of the shallow etched region in the purple channel. (d) Self-consistent band bending calculation at different gate potentials for $V_G$ = 0 V and -0.2 V. The dot-dashed line is the electron Fermi level. (e) Source-drain current in B1-L1 channel as a function of the gate voltage G for different source-drain bias voltages at 1.6 K, and $I_{sd}$ versus $V_{sd}$ at $V_G$ = -1.5 V is plotted in the inset.}
    \label{fig:3}
\end{figure}

In order to realize independent contact for the GaSb QW, a different procedure is applied, which utilizes a shallow etching of the InAs QW to prevent a drastic change of band bending in GaSb. Fig.~\ref{fig:3}(a) shows a sketch of the studied channel in the device with contacts S and D used for independently contacting GaSb QW. This was done by etching trenches from the top surface of the structure to deplete the upper InAs QW. It should be emphasized that once the upper InAs QW is removed completely, the surface Fermi level becomes pinned on AlSb and shifted upward by half an electron volt. This will result in a drastic change of band bending in GaSb QW, and there will be no free carriers in the GaSb QW even by applying a negative bias top gate potential.  Fig.~\ref{fig:3}(d) shows a band-structure design, which is capped with a 2-3 nm thick InAs remaining layer. It can be found that due to the surface Fermi level pinning of the remaining InAs layer, a small negative bias gate potential enables the creation of a conducting hole channel in the buried GaSb quantum well. Fig.~\ref{fig:3}(b) shows a cross-sectional profile of the shallow etched trench. After selectively etching down to the InAs QW,  we use a highly diluted solution of $\rm H_3PO_4/C_6H_8O_7/H_2O_2/H_2O$ to etch the InAs layer, which has a low etching rate and acts basically as a "polishing etch" in the case of shallow etching. Then, a dielectric layer of 30 nm-thick $\rm Al_2O_3$ is deposited, followed by evaporating TiAu, serving as a top gate. 

The optical image of the studied device is shown in Fig. 3(c). The purple channel corresponds to the sketch in Fig. 3(a), i.e., electrodes B1, L1 and G as the source, drain and control gate, respectively. In this device, electrodes L1, L2 and B1 are used to contact the InAs and GaSb QWs simultaneously by evaporating TiAu with annealing; Electrodes U1-U3 are fabricated as previously described by opening the dielectric window for separately attaching to InAs QW surface.  The right panel shows the zoomed optical image, where we can observe a shallow etched trench of about 4 $\rm \mu$m$\times$30 $\rm \mu$m size beneath the metal gate within the red box. The cross-sectional TEM image of the trench is shown in Fig. 3(b). Fig.~\ref{fig:3}(e) depicts the source-drain current in the B1-L1 channel as a function of the applied gate voltage G for fixed bias voltages at 1.6 K. The source-drain current through the lower GaSb QW can be controlled by the voltage applied to the top gate. When the gate voltage is applied negatively between 0 V and -0.2 V, the upper and lower QWs are both pinched off. As a sweep of the gate voltage to lower negative values, a conducting hole channel in GaSb QW is turned on, leading to an increase in the source-drain current. A similar behavior can be found in the B1-L2 channel. We plot the maximum saturation current $I_{max}$  at $V_G$ = -1.5 V versus the bias voltage $V_{sd}$ in the inset. A linear increase of source-drain current with $V_{sd}$ at fixed gate voltages is observed, indicating an ohmic contact realized by the above-described process to the lower GaSb QW. We comment on this: Although the severe surface scattering in the etching region may result in a high resistance level of the GaSb channel in the open state in Fig. 3(e), its good ohmic linearity ensures the effectiveness of the independent contact to GaSb QW.

\begin{figure}[!htb]
	\begin{center}
		\includegraphics[width=0.8\linewidth]{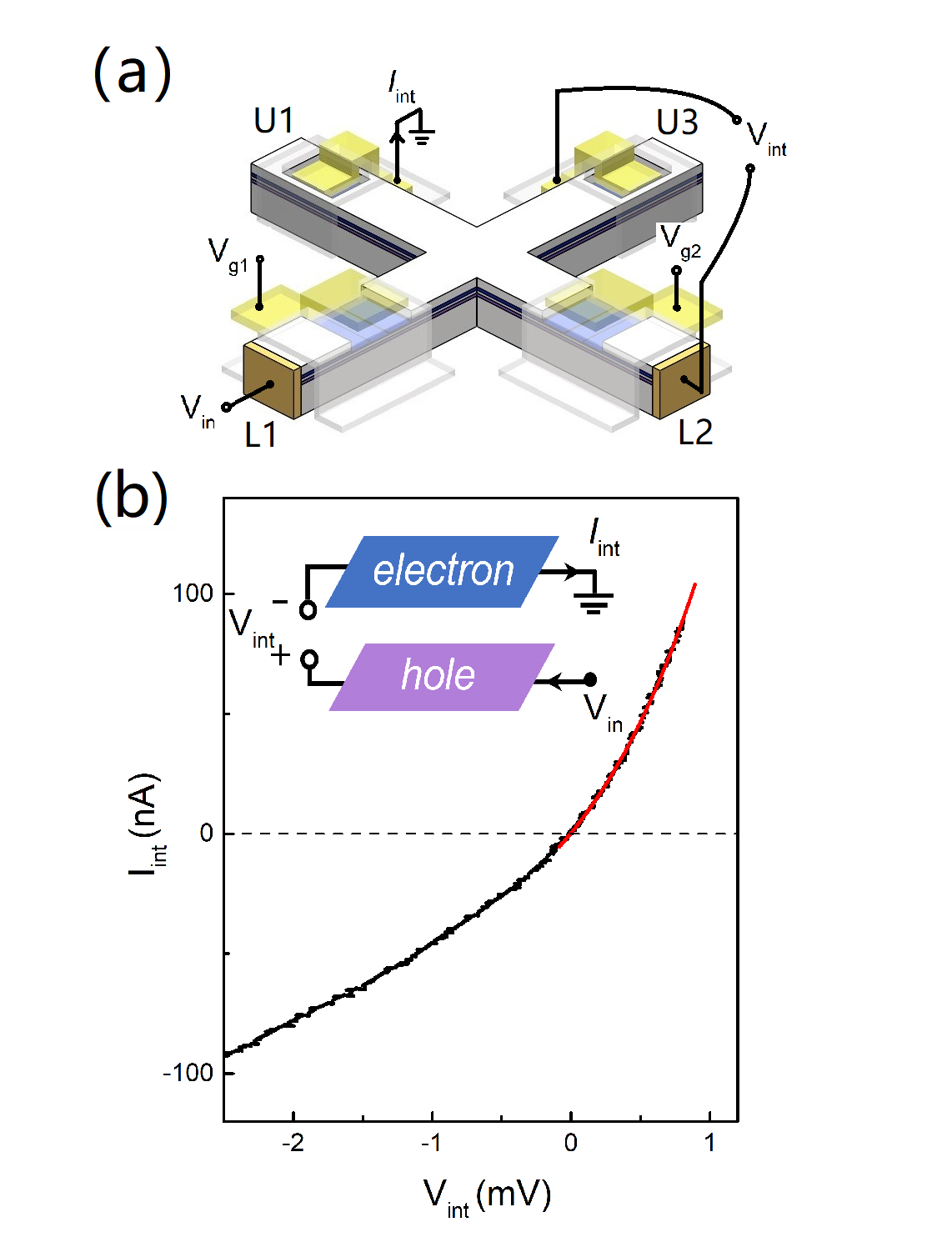}
	\end{center}
    \caption{Interlayer current-voltage characteristic in barrier-separated InAs/GaSb. (a) Sketch of the four-terminal section of the device with contacts L1,L2 for GaSb QW and contacts U1,U3 for InAs QW. (b) Nonlinear interlayer I-V characteristic when GaSb QW is positively biased and InAs QW negatively biased at $V_{g1} = V_{g2} $ = -1.2 V. The schematic representation of the measurement circuit is shown in the inset. }
    \label{fig:4}
\end{figure}

Figure ~\ref{fig:4}(a) shows a sketch of the four-terminal section of the device in Fig. 3, where contacts U1, U3 are attached to the upper InAs QW and contacts L1, L2 to the lower GaSb QW. We use a diamond pencil to cut off the arm of contact B1 in Fig. 3(c), so that there are no contacts to short the two layers. The interlayer current-voltage characteristic curve is measured with the setup in the inset of Fig. 4(b). We find that the interlayer isolation is not ideal in this sample, thus a certain degree of interlayer hybridization may actually occur. Many extrinsic causes may prevent the AlSb barrier from being completely insulating. e.g., defects incorporated during the AlSb growth\cite{aaberg2008intrinsic,erhart2010extrinsic} and the oxidization of AlSb when exposed to the air after sample growth\cite{wu2003sb}. Interestingly, the interlayer $I$-$V$ curve shows a nonlinear behavior of diode-like characteristics. Under forward bias, i.e., applying a positive voltage to the GaSb QW and a negative voltage to the InAs, the InAs/AlSb/GaSb junction carriers more current than under reverse bias, and approximately, the current increases exponentially (see the red line with an exponential fit). So far a standard p-n junction model seems unable to fully explain the observed $I$-$V$ curve under reverse bias. We are able to rule out contact-related issues and verify that the nonlinearity is intrinsically associated with the InAs/AlSb/GaSb junction (see supplementary material) where interlayer hybridization could be an important cause of nonlinear behavior. More work is required to further understand the interlayer nonlinear behavior and its dependence on hybridization, which will be the topic for future study. 

In summary, we have demonstrated a technique to produce separate ohmic contacts to closely spaced 2D electron-hole gas layers. By utilizing the surface Fermi level pinning position in InAs, independent contacts to barrier-separated InAs/GaSb systems have been realized, thus enabling the implementation of many experiments such as Coulomb drag, counterflow transport, and resonant interlayer tunneling studies. The capability of separately probing the interlayer-correlated states paves the way for studying the rich phase diagram of strongly interacting particles in Coulomb and tunnel-coupled InAs/GaSb systems.

\section*{Supplementary Material}
See supplementary material for more details about the basic material quality metrics of the sample, magnetoresistance oscillation of the 2DHG at a fixed magnetic field and interlayer nonlinearity.

\begin{acknowledgments}
We thank Gerard Sullivan for preparing the InAs/GaSb quantum wells by molecular beam epitaxy. The work at Beijing Academy of Quantum Information Sciences was financially supported by National Natural Science Foundation of China under Grant No. 12004039 and by Natural Science Foundation of Ningbo (Grant No. 2019A610068). The work at Peking University was supported by the National Key Research and Development Program of China (Grant No. 2019YFA0308400), and by the Strategic Priority Research Program of the Chinese Academy of Sciences (Grant No. XDB28000000).
\end{acknowledgments}

\nocite{*}
\bibliography{reference}
\bibliographystyle{aipnum4-1}
\end{document}